\documentclass[11pt,twoside]{article}

\usepackage{asp2004}
\usepackage{epsf}
\usepackage{psfig}
\usepackage{lscape}

\begin{document}
\title{Aromatic Features in M101 HII Regions and Starburst Galaxies}   
\author{Karl D.\ Gordon, Charles Engelbracht, J.D.T. Smith, George Rieke, 
  and Karl Misselt}   
\affil{Steward Observatory, Univ.\ of Arizona, Tucson, AZ 85721}    

\begin{abstract} 
The behavior of the aromatic features as a function of
metallicity/radiation field hardness in M101 HII regions and starburst
galaxies is investigated using Spitzer/IRS spectra.  These spectra
reveal that the 6.2, 7.8+8.6, and 11.3~\micron\ aromatics have
different dependencies on metallicity/radiation field hardness.
Specifically, the 6.2 and 7.8+8.6~\micron\ aromatics are weak or
absent at a metallicity, 12+log(O/H), of $\sim$8.0 while the
11.3~\micron\ feature remains relatively strong.  These results apply
to both the M101 HII regions and starbursts showing that HII regions
can indeed be thought of as ``mini-starbursts.''  Comparison of this
work with various candidate materials results in a good match with
the annealing behavior in Quenched Carbonaceous Composite (QCC) which
is an amorphous, hydrogenated, carbonaceous solid.
\end{abstract}



The aromatic features are a family of dust emission features seen in
many dusty astrophysical environments.  The strongest aromatics are
seen at 3.3, 6.2, 7.7, 8.6, 11.3, 12.7, and 17.1~\micron\ and all the
aromatics have been identified with C-H and C-C bending and stretching
modes of hydrocarbons containing aromatic rings \citep{Tielens05}.
The shape and strength of the aromatics have been seen to vary in
single sources
\citep{Werner04NGC7023}, among various Galactic sources
\citep{vanDiedenhoven04}, and among galaxies \citep{Engelbracht05SB,
Madden06, Wu06}.  A number of different materials have been proposed
as the carriers of these features since they were first discovered
\citep{Gillett73} including Hydrogenated Amorphous Carbon
\citep[HAC,][]{Duley83}, Quenched Carbonaceous Composites
\citep[QCC,][]{Sakata84}, Polycyclic Aromatic Hydrocarbons
\citep[PAHs,][]{Allamandola85}, coal \citep{Papoular89}, and
nanodiamonds \citep{Jones00}.  The leading candidate material for the
aromatics is PAH molecules and they are included in the modified
``astronomical PAHs'' form in dust grain models \citep{Li01, Zubko04}.
The preception that PAHs are the carrier of the aromatics can be seen
by the common use of the term ``PAH features'' to refer to the
aromatics.  The identification of PAH molecules with the aromatics may
be somewhat premature given that laboratory spectrum of a mix of PAHs
does not match the observed wavelengths and strengths of the aromatics
in detail
\citep{Hudgins04}.

\begin{figure}[!tb]
\plotfiddle{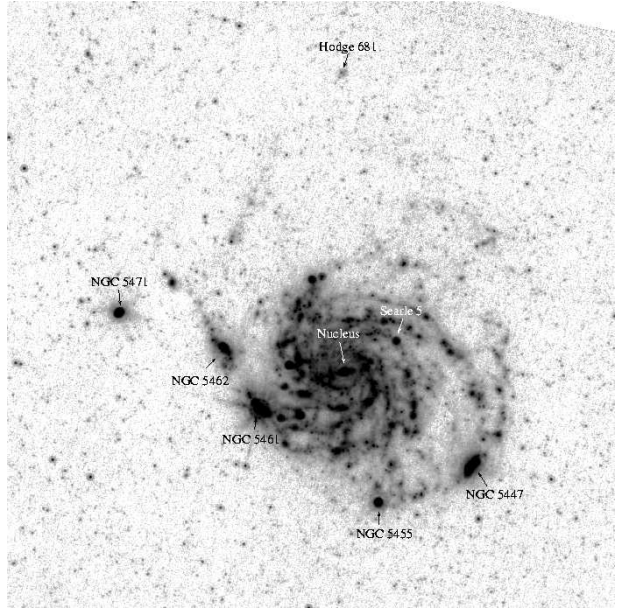}{3.0in}{0}{40}{40}{-110}{0}
\caption{The HII regions with IRS spectra are identified on the MIPS
24~\micron\ image.  The HII region Hodge 681 is
also identified (but not observed with IRS) as one of the lowest
metallicity HII region detected in the MIPS images.
\label{fig_m101_image}}
\end{figure}

As part of the Multiband Imaging Photometer for Spitzer (MIPS)
Guaranteed Time Observations (GTO), two programs were 
carried out to explore the variations in the infrared (IR)
Spectral Energy Distributions (SEDs) of massive star formation as a function
of metallicity.  The first program targets starburst galaxies
with metallicities [12+log(O/H)] between 7.2 to 8.9, with emphasis on
the lowest metallicity starbursts \citep{Engelbracht05SB}.  The second
program targets M101 as it is a large, face-on spiral galaxy with HII
regions which have metallicities from 7.4 to 8.8 \citep{Zaritsky94,
vanZee98, Kennicutt03}.  The combination of these two programs
probe HII regions, the components of starbursts, and entire starbursts.

Understanding the IR SEDs of starbursts and HII regions is important
to understanding dust properties as well as the SEDs of redshifted
galaxies undergoing massive star formation.  Probing how dust emission
features vary with physical environment (e.g., metallicity) can give
clues to the dust grain materials. Interpreting number counts in deep
fields to probe the star formation history of the Universe requires a
good understanding of galaxy SEDs; the behavior of the aromatics
directly affects models of number counts.  For example, the aromatics
moves through the MIPS 24~\micron\ band at redshifts between 1 and 2,
causing a peak in the number counts \citep{Lagache04}.

\section*{Observations}

The starburst and M101 HII regions have been observed with all the
instruments on the Spitzer Space Telescope
\citep[Spitzer][]{Werner04Spitzer}.  This paper concentrates on
preliminary results from the Infrared Spectrograph
\citep[IRS,][]{Houck04IRS}.  Preliminary results for M101 from the
Infrared Array Camera \citep[IRAC,][]{Fazio04IRAC} and MIPS
\citep{Rieke04MIPS} images were given by \citet{Gordon05NewViews}.
Preliminary results from the IRAC and MIPS imaging of the starbursts
were given by \citet{Engelbracht05SB}.

\begin{figure}[!tb]
\plotfiddle{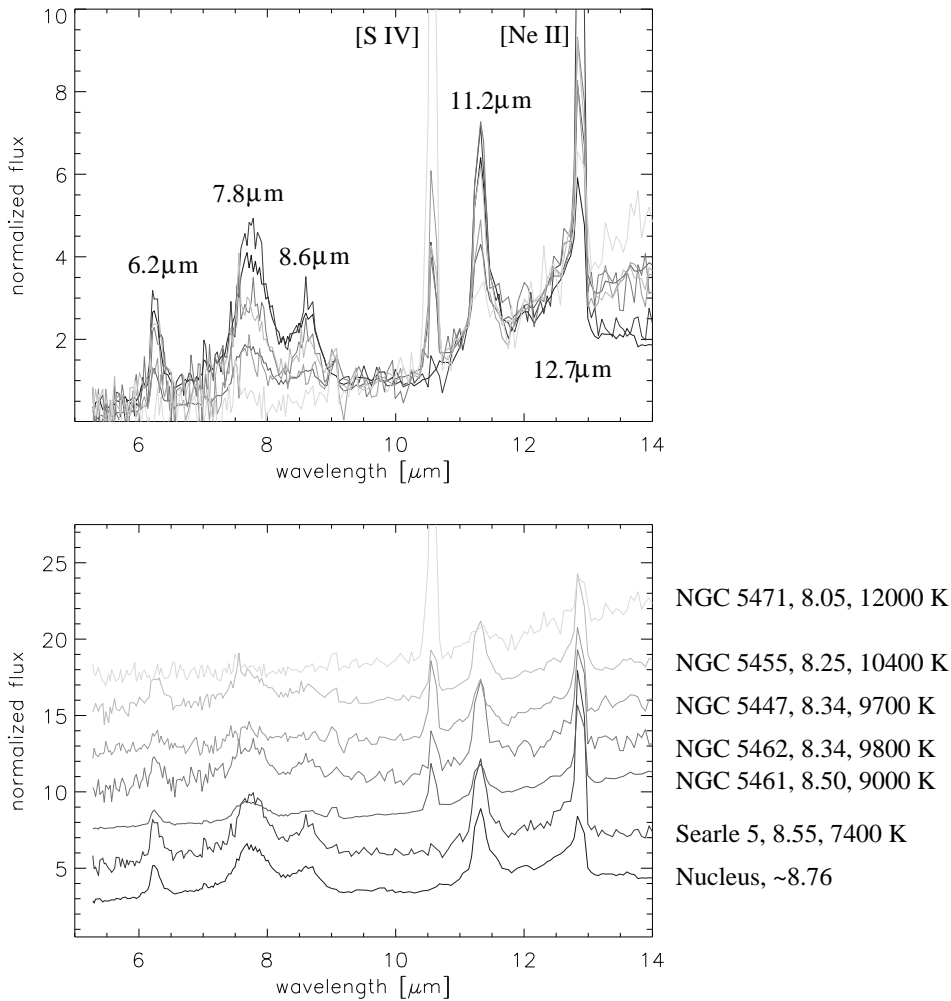}{5.05in}{0}{90}{90}{-200}{0}
\caption{IRS spectra of M101 HII regions are shown normalized at
10~\micron\ in the top plot and the same spectra are shown in the
bottom plot offset by an arbitrary amount.  The full IRS spectra
extend from 5-38~\micron.  The HII region names are listed along with
their measured metallicities and excitation temperatures
\citep{Kennicutt03}.
\label{fig_m101_spec}}
\end{figure}

The spectra of the M101 HII regions (Fig.~\ref{fig_m101_image}) were
taken in the spectral mapping mode 
with full slit width steps perpendicular to the slit.  Because the
ShortLow and LongLow observations were taken approximately 3 months apart, 
the orientations of the slits were approximately the same.  These
spectral mapping observations were reduced with the SSC pipeline
(version S12.0.2) and combined using CUBISM (Smith, J.D.T. 2006, in
prep.).  Spectra were extracted from the cubes for each HII region
using an aperture with a radius of 15\arcsec\ and sky
annuli from 20 to 30\arcsec\ referenced at 24~\micron.  The size of
the aperture and sky annuli were varied linearly with wavelength to
roughly account for the changing diffraction limited PSF.  The
resulting spectra are shown in Fig.~\ref{fig_m101_spec}.

\begin{figure}[!tb]
\plotfiddle{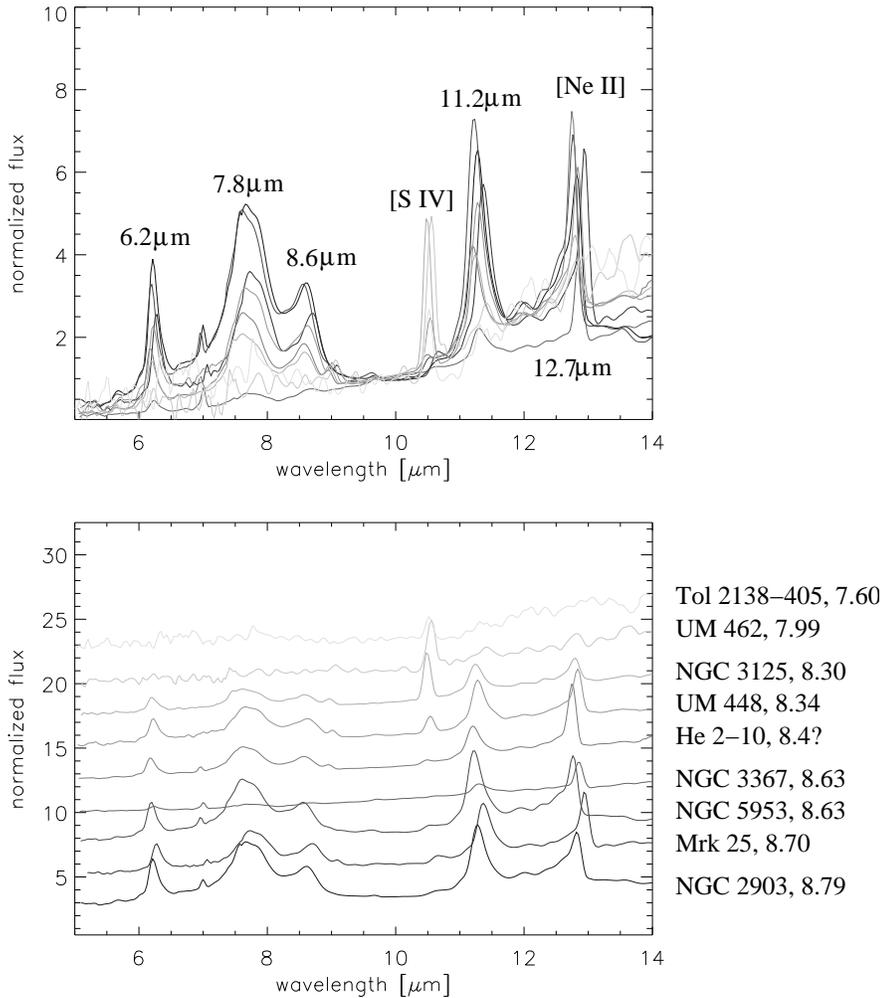}{5.05in}{0}{90}{90}{-200}{0}
\caption{IRS spectra of starburst galaxies are shown normalized at
10~\micron\ in the top plot and the same spectra are shown in the
bottom plot offset by an arbitrary amount.  The full IRS spectra
extend from 5-38~\micron.  The galaxy names are listed along with
their measured metallicities (Moustakas, J., in prep.).
\label{fig_sb_spec}}
\end{figure}

The spectra of the starburst galaxies were taken in staring mode.  The
reductions were done using the SSC pipeline (version S11.0.2).  The
redundant observations 
were combined and the spectra were extracted with a combination of
SPICE and IRAF. The extractions were for the full slit, since the galaxies
were usually noticeably extended. The final spectra have been
re-sampled to a common wavelength scale resulting in
some degree of smoothing.  The
resulting spectra are shown in Fig.~\ref{fig_sb_spec}.

\section*{Results}

The M101 HII region and starburst spectra shown in
Figs.~\ref{fig_m101_spec} \& \ref{fig_sb_spec} clearly display the
behavior of decreasing strength of the aromatics with metallicity
which is easiest to see in the strong 7.8+8.6~\micron\ aromatic
complex.  This is expected from previous Infrared Space Observatory
\citep{Madden06} and Spitzer \citep{Houck04SBS, Engelbracht05SB, 
Wu06} observations.  The spectra presented here allow us to determine
how individual features vary as a function of metallicity.  From
visual inspection of the spectra, the 6.2, 7.8, and 8.6~\micron\
features are weak or absent around a metallicity of 8.0 as seen from
the strength of these features in NGC~5471 (Fig.~\ref{fig_m101_spec})
and UM~462 and Tol2138-405 (Fig.~\ref{fig_sb_spec}).  This is not the
case for the 11.3 and 12.7~\micron\ features which are seen in all the
spectra in both plots except for the lowest metallicity starburst
spectra which are too noisy.  Thus, different aromatic features seem
to display different behaviors with metallicity.

\begin{figure}[!tb]
\plottwo{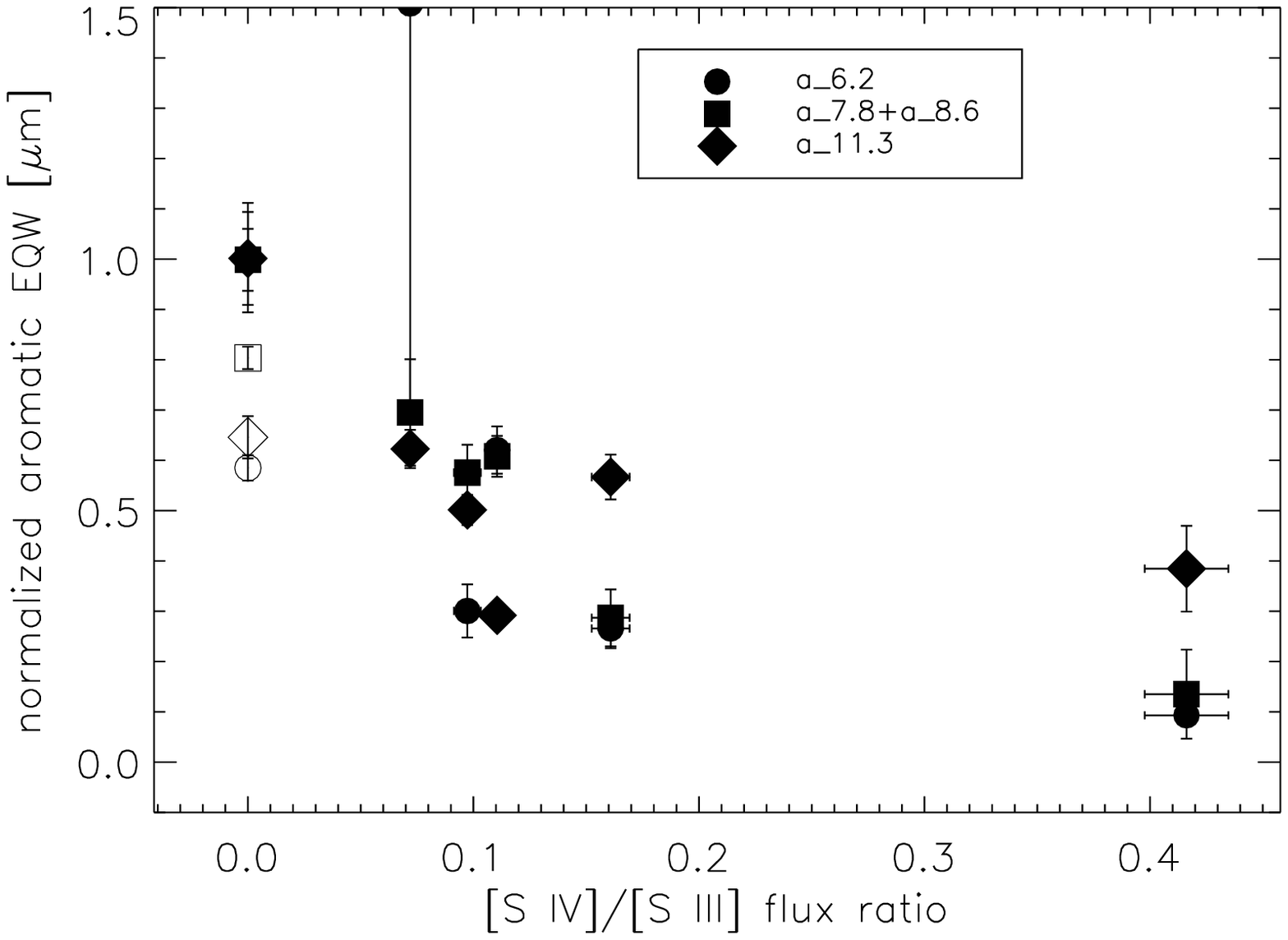}{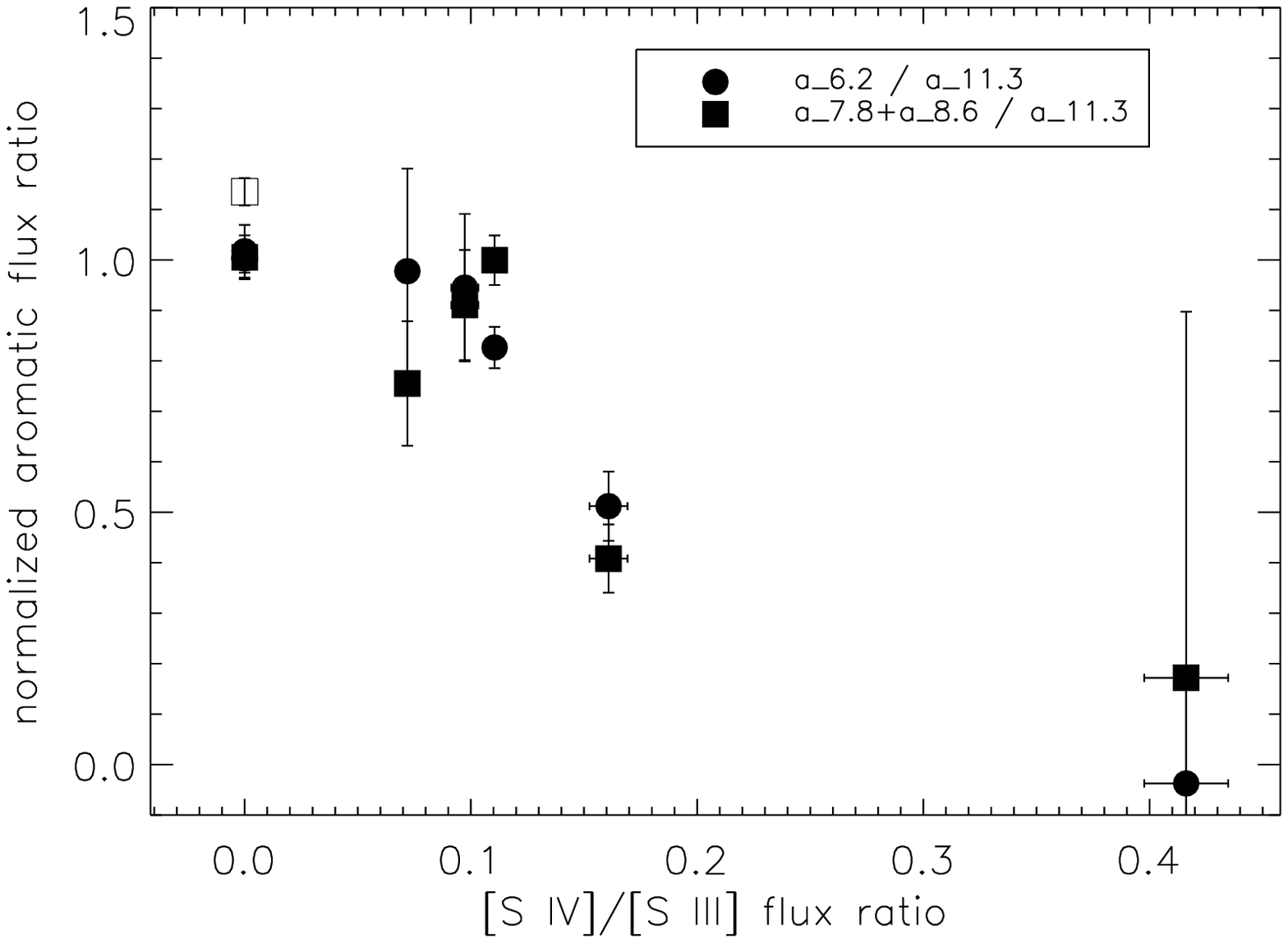} \\
\plottwo{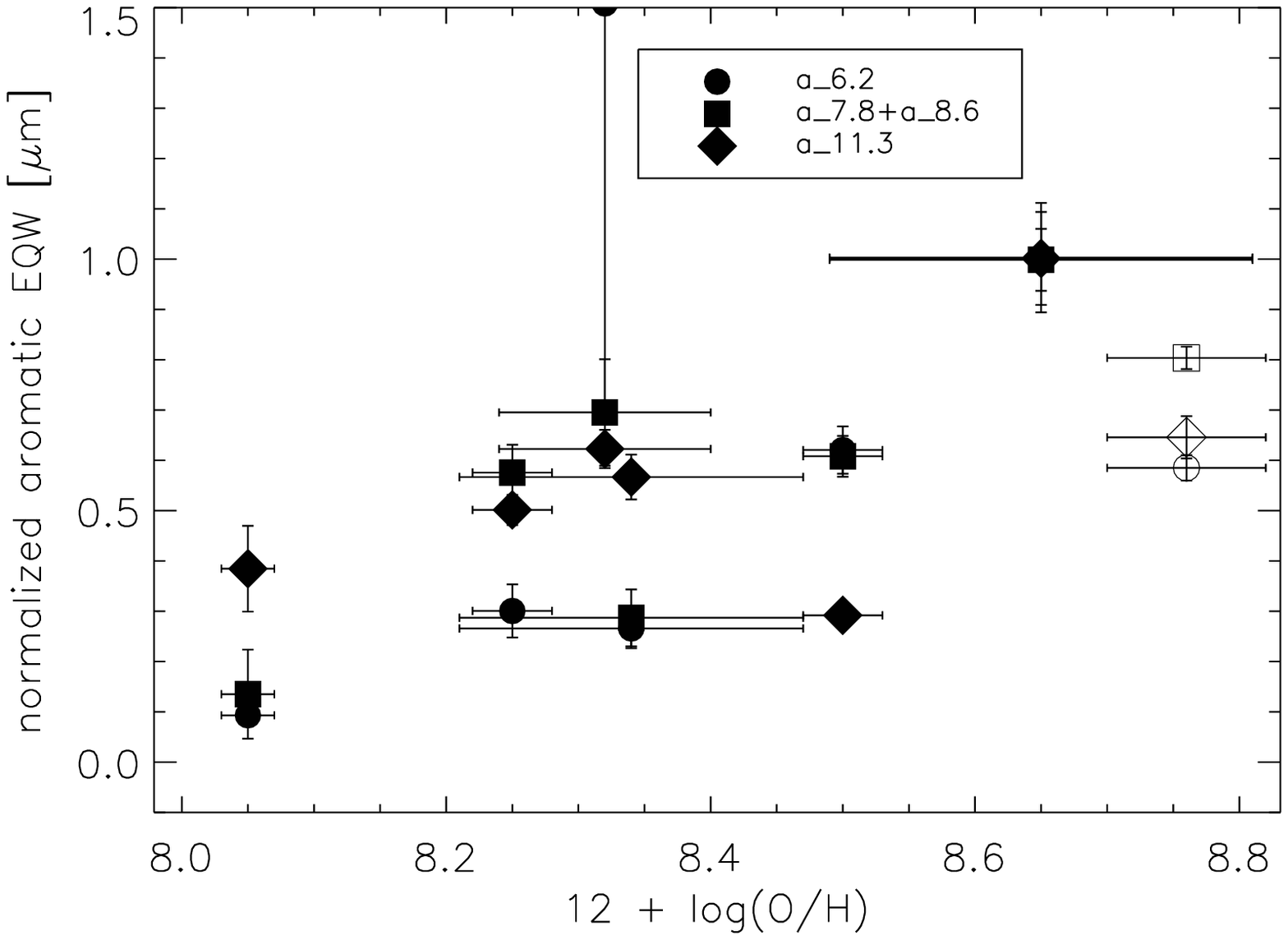}{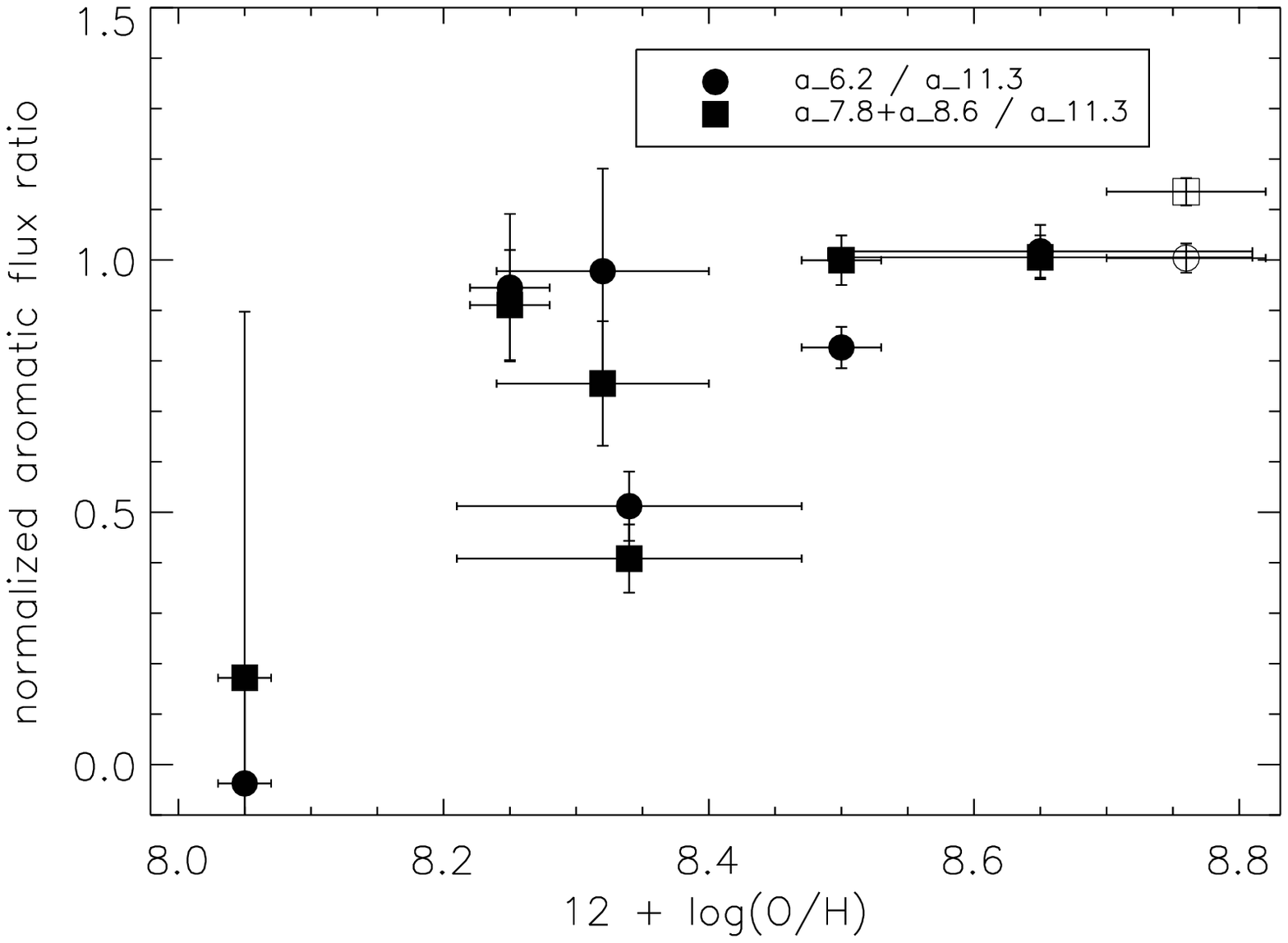}
\caption{The equivalent widths of 3 aromatics are plotted on the left
for the M101 HII regions.  The ratio of the aromatic to the
11.3~\micron\ aromatic is shown on the right.  The measurements for
the M101 nucleus are plotted with open symbols.
\label{fig_meas}}
\end{figure}

We have quantified the behavior of the aromatics in the M101 HII region
spectra by simply measuring the equivalent widths and total flux in
the 6.2, 7.8+8.6, and 
11.3 ~\micron\ aromatic features/complexes.  The measurements were done for
these three features/complexes as it is possible to identify continuum
on either side of them.  In addition to
these aromatics, we also measured the fluxes in the [S~IV]~10.6~\micron\
and [S~III]~18.7~\micron\ lines as the ratio of these two atomic
emission lines is a measure of the hardness of the radiation field.  

The equivalent widths and flux ratios of these aromatics are plotted in
Fig.~\ref{fig_meas} normalized to Searle~5, the strongest aromatic emitter,
versus the [S~IV]/[S~III] ratio and metallicity.  The weakening of the
aromatics 
seems better correlated with increasing radiation field hardness
([S~IV]/[S~III]) than metallicity.  The plots of the aromatics to the
11.3~\micron\ aromatic show that the 11.3~\micron\ aromatic weakens at a much
slower rate than the 6.2 and 7.8+8.6~\micron\ aromatics.  In addition, the
changing of the aromatic ratios also seems better correlated with radiation
field hardness than metallicity.

This preliminary analysis suggests that the weakening of the aromatics is
due to the hardening of the radiation field and that the aromatic spectrum
changes as the features weaken.  The pattern of aromatic spectrum change is
not in agreement with the predictions for PAH molecule ionization
\citep[Fig.~5]{Bakes01} which retain the 6.2+7.8+8.6~\micron\ complex at
higher ionizations, but predicts weak or no 11.3~\micron\ aromatic.
The aromatic spectrum change does seem to be in 
qualitative agreement with the annealing behavior of QCC
\citep[Fig.~2]{Sakata84}. 

A more complete analysis of the M101 HII regions will be
published by Gordon et al. (in prep.) including a more accurate
decomposition of the aromatics using the PAHFIT program (Smith, J.D.T., in
prep.).  A more complete analysis of the starburst spectra will be
published by Engelbracht et al. (in prep.).





\end{document}